\documentclass[preprint,groupedaddress,pre]{revtex4}
\usepackage{amssymb}
\usepackage{amsmath}

\input{tcilatex}

\begin{document}

\title{Stiff \ stability of the hydrogen atom in dissipative Fokker
electrodynamics}
\author{Jayme De Luca}
\email[author's email address:]{ deluca@df.ufscar.br}
\affiliation{Universidade Federal de S\~{a}o Carlos, \\
Departamento de F\'{\i}sica\\
Rodovia Washington Luis, km 235\\
Caixa Postal 676, S\~{a}o Carlos, S\~{a}o Paulo 13565-905}
\date{\today }

\begin{abstract}
We introduce an ad-hoc electrodynamics with advanced and retarded Li\'{e}%
nard-Wiechert interactions plus the dissipative Lorentz-Dirac
self-interaction force. We study the covariant dynamical system of the
electromagnetic two-body problem, i.e., the hydrogen atom. We perform the
linear stability analysis of circular orbits for oscillations perpendicular
to the orbital plane. In particular we study the normal modes of the
linearized dynamics that have an arbitrarily large imaginary eigenvalue.
These large eigenvalues are fast frequencies that introduce a fast (stiff)
timescale into the dynamics. As an application, we study the phenomenon of
resonant dissipation, i.e., a motion where both particles recoil together in
a drifting circular orbit (a bound state), while the atom dissipates
center-of-mass energy only. This balancing of the stiff dynamics is
established by the existence of a quartic resonant constant that locks the
dynamics to the neighborhood of the recoiling circular orbit. The resonance
condition quantizes the angular momenta in reasonable agreement with the
Bohr atom. The principal result is that the emission lines of quantum
electrodynamics (QED) agree with the prediction of our resonance condition
within one percent average deviation.
\end{abstract}

\pacs{05.45.-a, 41.20.-q, 02.30.Ks}
\maketitle

\section{Introduction}

In this paper we experiment with the stability analysis of the circular
orbits of the electromagnetic two-body problem. The main motivation is to
understand the complex dynamics described by the electromagnetic equations
of motion, that involve delay and third derivatives. We give a method to
derive the linearized equations of motion in the neighborhood of the
circular orbits of this implicitly-defined dynamical system with delay. We
introduce an ad-hoc electromagnetic-like setting that uses advanced and
retarded Li\'{e}nard-Wiechert interactions plus the dissipative
Lorentz-Dirac self-interaction force \cite{Dirac}, henceforth called the
dissipative Fokker setting (DFS). We study in detail a specific feature of
the tangent dynamics of the circular orbits of the two-body problem; The
stiff normal modes of the linearized dynamics, that have an arbitrarily
large imaginary eigenvalue. Last, we discuss an application for the hydrogen
atom and the surprising predictions of stability analysis within the DFS; We
predict several features of the Bohr atom \cite{Bohr} with high precision
and qualitative detail. A subset of the emission lines predicted by the DFS
agrees with the lines of quantum electrodynamics (QED) within one percent
average deviation. There is also a surprising body of qualitative agreement
with QED; \ (i) the emitted frequency is different from the orbital
frequency, (ii) the stable orbits of the DFS have angular momenta that are
multiples of a basic angular momentum. This basic angular momentum of the
DFS agrees well with Planck's constant and depends only logarithmically on
the mass of the heavier particle.

Dirac's 1938 fundamental work \cite{Dirac} on the electrodynamics of point
charges gave complex and stiff delay equations that were seldom studied.
Among the few models studied within Dirac's theory, the system of Eliezer's
theorem \cite{Eliezer, Parrott, Andrea, Massimo} revealed a surprising
dynamics; An electron moving in a Coulomb field with inclusion of
self-interaction can never fall into the center of force by radiating
energy. The result was generalized to motions in arbitrary attractive
potentials \cite{Parrott}, as well as to tridimensional motions with
self-interaction in a Coulomb field \cite{Andrea, Massimo}, finding that
only scattering states are possible. Since our model has Eliezer's problem
as the infinite-mass limit, a finite mass for the proton is essential for a
physically meaningful dynamics; If the proton has a finite mass, there is no
inertial frame where it rests at all times, and this in turn causes delay
because of the finite speed of light. It is widely known that QED gives a
satisfactory and precise description of atomic physics, but the same is not
popularly thought about atomic models based on classical electrodynamics.
Since dynamical studies are still missing, clearly this complex dynamics
needs to be investigated beyond our preliminary findings. Even though we are
not trying to replace QED, our understanding of this two-body dynamics might
prove useful for atomic physics and perhaps we can understand QED as the
effective theory of this complex stiff dynamics with delay. We shall
describe the two-body motion in terms of the familiar \emph{center-of-mass
coordinates} and \emph{coordinates of relative separation}, defined as the
familiar coordinate-transformation that maps the two-body Kepler problem
into the one-body problem with a reduced mass. We stress that in the present
relativistic motion the Cartesian center-of-mass vector is not ignorable,
and it represents three extra coupled degrees-of-freedom. We introduce the
concept of resonant dissipation to exploit this coupling and the many
solutions that a delay equation can have. Resonant dissipation is the
condition that both particles decelerate together, i.e., the center-of-mass
vector decelerates, while the coordinates of relative separation perform an
almost-circular orbit, despite of the energy losses of the metastable
center-of-mass dynamics.

Historically N\"{o}rdstrom \cite{Nordstrom, Page} suggested the use of
advanced and retarded potentials in atomic physics already in 1920, but the
self-interaction theory was problematic in 1920 and the idea disappeared.
The theory of nonlinear dynamics was not out yet in 1938 when Dirac's theory
for the electrodynamics of point charges appeared \cite{Dirac}, neither in
the Glorious days of the twentieth century physics \cite{Glorious}, such
that the our present experiment is a new application of modern nonlinear
dynamics. Advanced interactions appeared again in 1945, when Wheeler and
Feynman \cite{Fey-Whe,Narlikar} gave an electrodynamics based on the
postulate that every field is produced by charges located somewhere \cite%
{Mehra}. The theory was called action-at-a-distance electrodynamics \cite%
{Fey-Whe,Narlikar}, \ a theory where the isolated two-body problem is
defined by Fokker's action 
\begin{equation}
S_{F}=-\int m_{1}ds_{1}-\int m_{2}ds_{2}-e_{1}e_{2}\int \int \delta (||%
\mathbf{x}_{1}-\mathbf{x}_{2}||^{2})\mathbf{\dot{x}}_{1}\cdot \mathbf{\dot{x}%
}_{2}ds_{1}ds_{2},  \label{Fokker}
\end{equation}%
with $\mathbf{x}_{i}$ , $s_{i}$, $m_{i}$ and $e_{i}$ representing the
four-position, the proper time, the mass and the charge of particles $i=1,2$
respectively. In Eq. (\ref{Fokker}) the dot indicates the Minkowski scalar
product of four-vectors and double bars stand for the four-vector modulus 
\cite{Fey-Whe,Narlikar}. Due to the similarities with the equations of
motion of the DFS, the dynamical studies of the action-at-a-distance theory
are relevant for the present work. For example, in the collision of two
electrons with equations of motion determined by Eq. (\ref{Fokker}), the
solution is determined by initial position and velocity only, as proved in
Ref \cite{Drivergroup} (a Banach-to-Banach contraction mapping proof for
nonrunaway orbits). This suggests that we are dealing with a perfectly
causal and well-posed dynamical system dressed in unusual form \cite%
{wellposed}. Driver's result \cite{Drivergroup} suggests that a dynamics
with advance and delay is well-posed in the same way. The DFS presents
exactly the same neutral-delay mathematical problem of any
electromagnetic-like model, as for example the problem with retarded-only
fields of Refs.\cite{PRL,normalDeluca}. Fokker's action of Eq. (\ref{Fokker}%
) is used here to derive the sector of the DFS equations of motion
determined by the semi-sum of Li\'{e}nard-Wiechert fields. Last, advanced
interactions appeared again in another work of Eliezer; a generalization of
Dirac's covariant subtraction of electromagnetic infinities\cite%
{EliezerReview}. The resulting generalized electromagnetic settings include
the advanced interactions naturally, and provide a testbed for future
studies in electrodynamics \cite{EliezerReview}. Here we shall keep to the
DFS as a generic electromagnetic-like example.

The road map for this paper is as follows; In Section IV we give the main
technical part of the paper; We outline an economical method to derive the
tangent dynamics of the circular orbit based on a quadratic expansion of the
implicit light-cone condition. In this Section we also take the stiff limit
of the linear modes of the tangent dynamics. In Section V we give an
application to atomic physics, by discussing a necessary condition for the
state of resonant dissipation; This condition is heuristically expressed by
a simple resonance condition that predicts the correct atomic scales. The
earlier sections are a prelude to Section IV. In Appendix A we discuss how
the DFS can be fit into Dirac's electrodynamics of point charges. Section II
is a review of the circular orbit solution, to be used in Section IV and in
Section V. In Section III we build familiarity with Fokker's action of the
action-at-a-distance electrodynamics as a prelude to the quadratic
expansions needed for the linear stability analysis of Section IV. Last, in
Appendix B we discuss the soft normal modes of the tangent dynamics and in
Section VI we put the conclusions and discussion.

\bigskip

\bigskip

\section{The circular orbit solution}

\bigskip

In this Section we review the circular-orbit solution of the isolated
electromagnetic two-body problem of the action-at-a-distance electrodynamics 
\cite{Schonberg,Schild}, to be used as the unperturbed orbit. For the
isolated electromagnetic two-body problem, the tangent dynamics studied in
the next section is straight Lyapunov stability analysis. In the DFS there
is also a very small force along the orbital plane of the circular orbit,
such that a non-drifting circular orbit is not a solution of the equations
of motion. In the DFS the tangent dynamics is the starting point of a
perturbation scheme to impose that the resulting dynamics is a drifting
circular orbit (the state of resonant dissipation).

We use the index $i=1$ for the electron and $i=2$ for the proton, with
masses $m_{1}$ and $m_{2}$ respectively, as of Eq. (\ref{Fokker}). We
henceforth use units where the speed of light is $c=1$ and $%
e_{1}=-e_{2}\equiv -1$ (the electronic charge). The circular orbit is
illustrated in Fig. 1; A motion of the two particles in concentric circles
with the same constant angular speed and along a diameter. This dynamics
satisfies the time-symmetric problem of Fokker's action (\ref{Fokker})
because the symmetric contributions from future and past generate a
resulting force normal to the velocity of each particle \cite%
{Schonberg,Schild}. The details of \ this relativistic orbit will be given
now; The constant angular velocity is indicated by $\Omega $, the distance
between the particles in light-cone is $r_{b}$ and $\theta \equiv \Omega
r_{b}$ is the angle that one particle turns while the light emanating from
the other particle reaches it (the light-cone time lag). The angle $\theta $
is the natural independent parameter of this relativistic problem. Each
particle travels a circular orbit with radius and scalar velocity defined by 
\begin{eqnarray}
r_{1} &\equiv &b_{1}r_{b},  \label{defradius} \\
r_{2} &\equiv &b_{2}r_{b},  \notag
\end{eqnarray}%
and 
\begin{eqnarray}
v_{1} &=&\Omega r_{1}=\theta b_{1},  \label{defvelocity} \\
v_{2} &=&\Omega r_{2}=\theta b_{2},  \notag
\end{eqnarray}%
for the electron and for the proton, respectively. The condition that the
other particle turns an angle $\theta $ during the light-cone time lag is 
\cite{Schild}%
\begin{equation}
b_{1}^{2}+b_{2}^{2}+2b_{1}b_{2}\cos (\theta )=1,  \label{circularcone}
\end{equation}%
and is henceforth called the unperturbed light-cone condition. In Appendix B
we calculate $b_{1\text{ }}$and $b_{2}$ in a power series of $\theta $ up to
the fourth order. Last, because of the rotational invariance of Fokker's
action, there is a conserved angular momentum perpendicular to the plane of
the orbit, that is evaluated in Ref. \cite{Schild} to be%
\begin{equation}
l_{z}=\frac{1+v_{1}v_{2}\cos (\theta )}{\theta +v_{1}v_{2}\sin (\theta )},
\label{angular-momentum}
\end{equation}%
where the units of $l_{z}$ are $e^{2}/c$, just that we are using a unit
system where $e^{2}=c=1$ \cite{Schild}. Equation (\ref{circularcone})
restricts $b_{1}$ and $b_{2}$ to be less than one such that for small values
of $\theta $ the angular momentum of Eq. (\ref{angular-momentum}) is of the
order of $l_{z}\sim \theta ^{-1}$. For orbits in the atomic magnitude, $%
l_{z}\simeq \theta ^{-1}$ is about one over the fine-structure constant, $%
\alpha ^{-1}=137.036$. It is curious to notice that each (advance/retarded)
interaction term of Fokker's action, Eq. (\ref{VAC} ), evaluates exactly to $%
\frac{1}{2}\Omega l_{z}$ along a circular orbit, with $l_{z}$ given by Eq. (%
\ref{angular-momentum}). This combination of angular momentum times the
orbital frequency is reminiscent of the formal maneuvers of quantum
mechanics.

\section{\protect\bigskip\ Fokker's Action}

We use Fokker's action in this work as a means to derive the sector of the
DFS equations of motion determined by the semi-sum of the Li\'{e}%
nard-Wiechert potentials. In the following we discuss the Lagrangian
formalism of Fokker's action (\ref{Fokker}) as an introduction to our
economical method to obtain the tangent dynamics by expanding this action to
quadratic order. The delta-function of Fokker's action (\ref{Fokker})
contains the retarded and the advanced light-cone contributions, and it is
convenient to separate those two parts by factoring the argument of the
delta function as%
\begin{equation}
(t_{1}-t_{2})^{2}-r_{12}^{2}=[t_{1}-t_{2}-r_{12}][t_{1}-t_{2}+r_{12}],
\label{splitlightcone}
\end{equation}%
where $r_{12}$ stands for the Cartesian distance between particle $1$ at
time $t_{1\text{ }}$and particle $2$ at time $t_{2\text{ }}$and each factor
on the right-hand side of Eq. (\ref{splitlightcone}) is related to the
advanced and the retarded light-cones of particle $1$ respectively. The
delta-function of a product argument is a sum of two delta-functions each
multiplied by the respective Jacobian, such that the interaction term of \
Fokker's action (\ref{Fokker}) can be written as 
\begin{eqnarray}
VA &=&\int \frac{1}{2r_{12}}\delta (t_{1}-t_{2}-r_{12})(1-\mathbf{v}%
_{1}\cdot \mathbf{v}_{2})dt_{1}dt_{2}  \label{VA} \\
&&+\int \frac{1}{2r_{12}}\delta (t_{1}-t_{2}+r_{12})(1-\mathbf{v}_{1}\cdot 
\mathbf{v}_{2})dt_{1}dt_{2},  \notag
\end{eqnarray}%
where $\mathbf{v}_{1}$ henceforth stands for the Cartesian velocity of
particle $1$ at time $t_{1}$ and $\mathbf{v}_{2}$ henceforth stands for the
Cartesian velocity of particle $2$ at time $t_{2}$. We henceforth use the
dot to indicate the scalar product of two Cartesian vectors, as already used
in Eq. (\ref{VA}). Integration of each term of Eq. (\ref{VA}) over $t_{2}$
brings out another Jacobian factor and yields 
\begin{equation}
VA=\int \frac{1}{2}\frac{(1-\mathbf{v}_{1}\cdot \mathbf{v}_{2a})}{r_{12}(1+%
\mathbf{n}_{12a}\cdot \mathbf{v}_{2a})}dt_{1}+\int \frac{1}{2}\frac{(1-%
\mathbf{v}_{1}\cdot \mathbf{v}_{2b})}{r_{12}(1-\mathbf{n}_{12b}\cdot \mathbf{%
v}_{2b})}dt_{1},  \label{VAintegr}
\end{equation}%
where $\mathbf{n}_{12a}$ is a unit vector connecting the advanced position
of particle $2$ at time $t_{2}$ to the position of particle $1$ at time $%
t_{1}$,  vector $\mathbf{n}_{12b}$ is a unit vector connecting the retarded
position of particle $2$ at time $t_{2}$ to the position of particle $1$ at
time $t_{1}$and\textbf{\ }$\mathbf{v}_{2a}$ and $\mathbf{v}_{2b}$ stand for
the velocity of particle $2$ at the advanced and retarded time $t_{2}$
respectively. Equation (\ref{VAintegr}) is the most useful form of Fokker's
interaction for our purposes. Notice that each term of Eq. (\ref{VAintegr})
can be cast in the form%
\begin{equation}
\frac{1}{2}\int \frac{(1-\mathbf{v}_{1}\cdot \mathbf{v}_{2c})}{r_{12}(1+%
\frac{\mathbf{n}_{12}\cdot \mathbf{v}_{2c}}{c})}dt_{1}\equiv \frac{-1}{2}%
\int (V-\mathbf{v}_{1}\cdot \mathbf{A})dt,  \label{VAC}
\end{equation}%
where $V$ and $\mathbf{A}$ are the Li\'{e}nard-Wiechert scalar potential and
the Li\'{e}nard-Wiechert vector potential respectively. We have introduced
the quantity $c=\pm 1$ in the denominator of Eq. (\ref{VAC}) such that $c=1$
represents the advanced interaction while $c=-1$ represents the retarded
interaction. The quantities of particle $2$ in Eq. (\ref{VAC}) \ are to be
evaluated at a time $t_{2}$ defined implicitly by 
\begin{equation}
t_{2}=t_{1}+\frac{r_{12}}{c},  \label{lightcone}
\end{equation}%
where $c=\pm 1$ describes the advanced and retarded light cones,
respectively. Because of this decomposition of Fokker's interaction into $V$
and $\mathbf{A}$ parts, we henceforth call Eq. (\ref{VAC}) \ the VA
interaction. A derivation of the Li\'{e}nard-Wiechert potentials from
Fokker's action and details such as the Darwin approximation are found in
Ref.\cite{Anderson}.

The stiff limit is determined by the largest-order derivative appearing in
the linearized equations of motion of Appendix A. In this approximation, the
contribution of the self-interaction force to the linearized dynamics about
a circular orbit is simply given by the Abraham-Lorentz -Dirac force%
\begin{equation}
\mathbf{F}_{rad}=\frac{2}{3}\mathbf{\dot{a}}\text{.}  \label{LDE}
\end{equation}%
The contribution of the other smaller terms will be given elsewhere.

\section{\protect\bigskip Linear stability analysis}

In this Section we study the linear stability analysis of the circular
orbits for displacements perpendicular to the orbital plane, henceforth
called the $z$-direction. We give an economical method to obtain these
equations of tangent dynamics by expanding the implicit light-cone condition
up to quadratic order. We start from the equations of motion of the isolated
system, which are derived from Fokker's action (\ref{Fokker}) and yield the
Li\'{e}nard-Wiechert fields in the half-retarded plus half-advanced
combination. This linearized $z$ dynamics is uncoupled from the planar
dynamics, and the linearized equations can de derived without the use of a
symbolic software, as we explain in the following. The Cartesian coordinates
of a transversely perturbed circular orbit are defined by 
\begin{eqnarray}
x_{k}+iy_{k} &\equiv &r_{b}d_{k}\exp (i\Omega t),  \label{Zperturb} \\
x_{k}-iy_{k} &\equiv &r_{b}d_{k}^{\ast }\exp (-i\Omega t),  \notag \\
z_{k} &\equiv &r_{b}CSZ_{k},  \notag
\end{eqnarray}%
where $k=1$ for the electron and $k=2$ \ for the proton, $Z_{k}$ is the
small transverse perturbation, $d_{1}\equiv b_{1}$ and $d_{2}\equiv -b_{2}$
are defined from the two real parameters of Eq. (\ref{defradius}) and $%
\Omega $ is the orbital frequency defined above Eq. (\ref{defradius}). Last,
in Eq. (\ref{Zperturb}) $C$ and $S$ are defined by 
\begin{equation}
C\equiv 1+b_{1}b_{2}\theta ^{2}\cos (\theta ),  \label{defC}
\end{equation}%
and%
\begin{equation}
S\equiv 1+b_{1}b_{2}\theta \sin (\theta ).  \label{defS}
\end{equation}

We henceforth introduce a scaled time $\tau \equiv \Omega t$. The linear
stability analysis involves expanding the equations of motion to linear
order in $Z_{k}$ , which in turn is determined by the quadratic expansion of
Fokker's action in $Z_{k}$. The main tool for expanding this quadratic
action is the perturbed light-cone condition, Eq. (\ref{lightcone}), about
the circular orbit (where $r_{12}=r_{b}$ is the constant circular lag). We
introduce a function $\varphi $ of the $Z_{1}$ and $Z_{2}$ perturbations by
expanding the light-cone time $t_{2}$ as 
\begin{equation}
t_{2}\equiv t_{1}+\frac{r_{b}}{c}+\frac{\varphi }{\Omega }.  \label{defphiZ}
\end{equation}%
In the following we calculate this homogeneous functional $\varphi $ of $%
Z_{1}$ and $Z_{2}$ up to quadratic order. The distance $r_{12}$ entering Eq.
(\ref{lightcone}) is to be evaluated from the position of particle $1$ at
time $t_{1}$, to the position of particle $2$ \ at the time $t_{2}$ defined
implicitly by Eq. (\ref{defphiZ}). The coordinates of particle $2$ at the
time $t_{2}$ are defined implicitly by 
\begin{eqnarray}
x_{2}+iy_{2} &=&r_{b}d_{2}\exp (i\tau _{1}+ic\theta +i\varphi ),
\label{Z2ret} \\
x_{2}-iy_{2} &=&r_{b}d_{2}^{\ast }\exp (-i\tau _{1}-ic\theta -i\varphi ), 
\notag \\
z_{2} &=&r_{b}CSZ_{2}(\tau _{1}+c\theta +\varphi )\equiv r_{b}CSZ_{2c}, 
\notag
\end{eqnarray}%
where $c=1$ for the advanced time $t_{2}$ and $c=-1$ for the retarded time $%
t_{2}$. Notice that \ Eq. (\ref{Z2ret})\ \ defines the coordinates
implicitly, because $\varphi $ is a function of the deviations $Z_{1}$ and $%
Z_{2}$. Even though $\theta $ is small in applications of atomic physics, we
stress that one should never expand in powers of $\ \theta $; the correct
infinitesimal quantity of the linear stability analysis is the size of the
deviations from circularity and their homogeneous functions such as $\varphi 
$ (expanding in $\theta $ produces the Darwin approximation \cite{Anderson}%
). This non-analyticity will become clear after we show that the logarithm
of $\theta $ appears. We therefore expand the advanced/retarded position $%
Z_{2c}$ of particle $2$ at the scaled time $\tau _{1}+c\theta +\varphi $ in
a Taylor series in $\varphi $ about the advanced/retarded position $\tau
_{1}+c\theta $. It turns out that only the zeroth-order term appears in the
action up to quadratic order. Because of this, the linearized equations
involve only a constant shift, a considerable simplification. Substituting $%
t_{2}$ of Eq. (\ref{defphiZ}) together with the position (\ref{Z2ret}) of
particle $2$ into Eq. (\ref{lightcone}) and using Pythagoras theorem for the
distance $r_{12}$ from particle $1$ at time $t_{1}$ to particle $2$ at time $%
t_{2}$ yields%
\begin{equation}
r_{12}^{2}=(r_{b}+r_{b}\frac{\varphi c}{\Omega r_{b}}%
)^{2}=r_{b}^{2}+r_{b}^{2}C^{2}S^{2}(Z_{1}-Z_{2c})^{2}.  \label{lightconeZ}
\end{equation}%
Notice that the $Z$ variations decouple from the planar variations because
there is no mixed linear term of $Z$ times a linear perturbation of the
planar coordinate in Eq. (\ref{lightconeZ}); These are naturally separated
by Pythagoras theorem. The planar perturbations enter in Eq. (\ref%
{lightconeZ}) as an added quadratic form, as given in the next section. It
is convenient to define another function $\Phi $ by $\varphi \equiv \theta
cCS\Phi $, such that Eq. (\ref{lightconeZ}) is a quadratic equation of $\Phi 
$ and the regular solution up to second order in $Z_{1}$ and $Z_{2c}$ is%
\begin{equation}
\Phi =\frac{CS}{2}(Z_{1}-Z_{2c})^{2}.  \label{PHIZ}
\end{equation}%
The coordinate $Z_{2}$ appears evaluated at the advanced/retarded time in
Eq. (\ref{PHIZ}), and to obtain the action up to quadratic terms it is
sufficient to keep the first term $Z_{2c}=Z_{2}(\tau _{1}+c\theta +\varphi ) 
$ $\simeq Z_{2}(\tau _{1}+c\theta )$. Using the $z-$perturbed orbit defined
by Eq. (\ref{Z2ret}) to calculate the numerator of the VA interaction of Eq.
(\ref{VAC}) yields%
\begin{equation}
(1-\mathbf{v}_{1}\cdot \mathbf{v}_{2c})=1+\theta ^{2}\cos (\theta
)b_{1}b_{2}-\theta ^{2}C^{2}S^{2}\dot{Z}_{1}\dot{Z}_{2c},  \label{h2Z}
\end{equation}%
and the denominator of the VA interaction of Eq. (\ref{VAC}) is%
\begin{equation}
r_{12}(1+\mathbf{n}_{12c}\cdot \mathbf{v}_{2c}/c)=r_{b}S[1+C\Phi +\theta
cC^{2}S(Z_{1}-Z_{2c})\dot{Z}_{2c}].  \label{h4Z}
\end{equation}%
Notice that the quadratic term $Z_{2c}\dot{Z}_{2c}$ on the right-hand side
of Eq. (\ref{h4Z}) can be dropped because it represents an exact Gauge that
does not affect the Euler-Lagrange equations of motion, such that 
\begin{equation}
r_{12}(1+\mathbf{n}_{12c}\cdot \mathbf{v}_{2c}/c)\approx r_{b}S[1+C\Phi
+\theta cC^{2}SZ_{1}\dot{Z}_{2c}],  \label{h4ZGauge}
\end{equation}%
where the equivalence sign $\approx $ henceforth means equivalent up to a
Gauge term of second order. Even if a quadratic Gauge term appears in the
denominator, in an expansion up to quadratic order it would still produce a
Gauge and therefore it can be dropped directly from the denominator. One
should be careful not to do this with linear Gauges, which appear only in
the planar stability analysis to be considered elsewhere. In this way, the
expansion up to second order of the VA interaction of Eq. (\ref{VAC}) is
simply%
\begin{equation}
VA\approx (\frac{C}{2r_{b}S})\{1-\theta ^{2}CS^{2}\dot{Z}_{1}\dot{Z}_{2c}-%
\frac{C^{2}S}{2}(Z_{1}-Z_{2c})^{2}-\theta cC^{2}SZ_{1}\dot{Z}_{2c}\}.
\label{VAZ}
\end{equation}%
Last, we need the kinetic energy along the $z$-perturbed circular orbit,
which \ we express in terms of $Z_{1}$ of definition (\ref{Zperturb}) as%
\begin{equation}
T_{1}=-m_{1}\sqrt{1-v_{1}^{2}}=-\frac{m_{1}}{\gamma _{1}}\sqrt{1-\gamma
_{1}^{2}C^{2}S^{2}\theta ^{2}\dot{Z}_{1}^{2}},  \label{kineticZ}
\end{equation}%
where the dot means derivative with respect to the scaled time $\tau $, $%
\gamma _{1}^{-1}\equiv \sqrt{1-v_{1}^{2}}$ , and we have used $\Omega
r_{b}=\theta $. The expansion of Eq. (\ref{kineticZ}) up to second order is%
\begin{equation}
T_{1}=(\frac{C}{r_{b}S})\{\frac{-r_{b}Sm_{1}}{C\gamma _{1}}+\frac{\epsilon
_{1}}{2}\dot{Z}_{1}^{2}+...\},  \label{expaE}
\end{equation}%
where $\epsilon _{1}$ $\equiv m_{1}r_{b}\gamma _{1}\theta ^{2}CS^{3}$ is
calculated with Eq. (\ref{radialEq1}) to be%
\begin{equation}
\epsilon _{1}\equiv \frac{C}{b_{1}}\{[C^{2}+\theta
^{2}S(S-1)](b_{1}+b_{2}\cos (\theta ))+S(\theta \sin (\theta )-\theta
^{2}\cos (\theta ))b_{2}\}.  \label{mass1}
\end{equation}

We are ready to derive the Euler-Lagrange equation of motion for particle $1$
of the isolated two-body problem using the quadratic Lagrangian%
\begin{equation}
L_{1}=T_{1}+VA_{c=1}+VA_{c=-1}.  \label{LagrangeZ1}
\end{equation}%
This equation of motion is%
\begin{equation}
\epsilon _{1}\ddot{Z}_{1}=-\frac{C^{2}S}{2}(2Z_{1}-Z_{2+}-Z_{2-})-\frac{%
\theta C^{2}S}{2}(\dot{Z}_{2+}-\dot{Z}_{2-})-\frac{\theta ^{2}CS^{2}}{2}(%
\ddot{Z}_{2+}+\ddot{Z}_{2-}).  \label{EQZ1}
\end{equation}%
Notice that the term on the left-hand side of Eq. (\ref{EQZ1}) can be
written as 
\begin{equation}
\epsilon _{1}\ddot{Z}_{1}=r_{b}^{2}S^{2}m_{1}\gamma _{1}\Omega ^{2}CS\ddot{Z}%
_{1}=r_{b}^{2}S^{2}\frac{dp_{z}}{dt},  \label{add1}
\end{equation}%
which is proportional to the force along the $z$-direction. According to the
prescription of the DFS, we shall add the following self-interaction term to
the right-hand side of Eq. (\ref{EQZ1})%
\begin{equation}
r_{b}^{2}S^{2}\mathbf{F}_{rad}=\frac{2}{3}CS^{3}\dddot{Z}_{1},
\label{dissi1}
\end{equation}%
where the triple dot means three derivatives with respect to the scaled time
and we have used Eq. (\ref{LDE}). The full linearized equation of motion for 
$Z_{1}$ is%
\begin{equation}
\epsilon _{1}\ddot{Z}_{1}=\frac{2}{3}CS^{3}\dddot{Z}_{1}-\frac{C^{2}S}{2}%
(2Z_{1}-Z_{2+}-Z_{2-})-\frac{\theta C^{2}S}{2}(\dot{Z}_{2+}-\dot{Z}_{2-})-%
\frac{\theta ^{2}CS^{2}}{2}(\ddot{Z}_{2+}+\ddot{Z}_{2-}).  \label{DFEQZ1}
\end{equation}%
The linearized equation for $Z_{2}$ is completely analogous and is obtained
by interchanging $Z_{1\text{ }}$by $Z_{2}$ and $\epsilon _{1}$ by $\epsilon
_{2}$ in Eq. (\ref{DFEQZ1}). The general solution of a linear delay equation
can be obtained by Laplace transform \cite{Bellman} and is a linear
combination of the following normal mode solutions. A normal mode solution
is obtained by substituting $Z_{1}=A\exp (p\tau )$ and $Z_{2}=B\exp (p\tau )$
into the two linearized equations, and requires the vanishing of the
following $2\times 2$ determinant%
\begin{equation}
\det Z\equiv \left\vert 
\begin{array}{cc}
C^{2}S+\epsilon _{1}p^{2}-\frac{2}{3}CS^{3}\theta ^{3}p^{3} & G(\theta ,p)
\\ 
G(\theta ,p) & C^{2}S+\epsilon _{2}p^{2}-\frac{2}{3}CS^{3}\theta ^{3}p^{3}%
\end{array}%
\right\vert ,  \label{detZ}
\end{equation}%
where $G(\theta ,p)\equiv (C^{2}S-CS^{2}\theta ^{2}p^{2})\cosh (p\theta
)+C^{2}Sp\theta \sinh (p\theta )$. Two kinds of limits are interesting for
the infinite-dimensional formal collection of normal modes of Eq. (\ref{detZ}%
); (i) the four soft Coulombian modes obtained by expanding Eq. (\ref{detZ})
in powers of $\theta $ for small values of $p$, as discussed in Appendix B,
and (ii) the stiff limit obtained when $p\theta $ is large, such that the
hyperbolic functions of the $G(\theta ,p)$ acquire a large magnitude \cite%
{astar2B}. In the following we use the zeroth-order term of the expansion
for $b_{1}$ and $b_{2}$ given in Appendix B to evaluate the determinant (\ref%
{detZ}) :\ 

\begin{eqnarray}
\epsilon _{1} &=&\frac{M}{m_{2}}+O(\theta ^{2}),  \label{zeroth} \\
\epsilon _{2} &=&\frac{M}{m_{1}}+O(\theta ^{2}),  \notag \\
C &=&1+O(\theta ^{2}),  \notag \\
S &=&1+O(\theta ^{2}).  \notag
\end{eqnarray}%
For small $\theta $, the second-order and higher even-order terms of \ Eq. (%
\ref{zeroth}) give only a small correction. Substituting Eq. (\ref{zeroth})
into Eq. (\ref{detZ}) and defining $p\equiv \lambda /\theta $, we obtain%
\begin{equation}
\frac{\mu \theta ^{4}}{M\lambda ^{4}}(\det Z)=1-\frac{2}{3}\theta
^{2}\lambda +\frac{4}{9}\frac{\mu }{M}\theta ^{4}\lambda ^{2}-\frac{\mu
\theta ^{4}}{M}[(1-\frac{1}{\lambda ^{2}})\cosh (\lambda )-\frac{1}{\lambda }%
\sinh (\lambda )]^{2},  \label{EQ5Z}
\end{equation}%
where we have dropped small $O(\theta ^{2})$ terms. The stiff-mode condition
defined by Eq. (\ref{detZ}) ($\det Z=0$) is%
\begin{equation}
1-\frac{2}{3}\theta ^{2}\lambda +\frac{4\mu }{9M}\theta ^{4}\lambda ^{2}-%
\frac{\mu \theta ^{4}}{M}[(1-\frac{1}{\lambda ^{2}}+\frac{1}{\lambda ^{4}}%
)\cosh ^{2}(\lambda )+\frac{1}{\lambda }(1-\frac{1}{\lambda ^{2}})\sinh
(2\lambda )]=0.  \label{fourthZ}
\end{equation}%
For future reference we give also the stiff limit for the $z$-tangent
dynamics without the self-interaction terms, which is obtained from Eq. (\ref%
{EQZ1}) and the corresponding equation for particle $2$%
\begin{equation}
1-\frac{\mu \theta ^{4}}{M}[(1-\frac{1}{\lambda ^{2}}+\frac{1}{\lambda ^{4}}%
)\cosh ^{2}(\lambda )+\frac{1}{\lambda }(1-\frac{1}{\lambda ^{2}})\sinh
(2\lambda )]=0.  \label{actionZ}
\end{equation}

\section{\protect\bigskip\ The stiff \ stability of the hydrogen atom}

We are interested in finding motions where the particles recoil together
while staying in the neighborhood of a drifting circular orbit, i.e., the
state of resonant dissipation. The need for a resonance becomes obvious in
the following perturbative scheme; (i) We take the circular orbit as the
unperturbed state. (ii) We substitute the circular orbit plus a perturbation
into the equations of motion of the DFS and take the linearized equations of
motion. The circular orbit is not an exact solution of the DFS equations of
motion, because of the small forcing coming from the third derivatives. This
perturbative scheme yields linear delay equations with a small forcing term
along the orbital plane. It is then possible to show by averaging \cite{to
be published} that a weakly-accelerated drifting circular orbit is never a
solution to these linear equations. Therefore, a bifurcation of the circular
orbit must happen and a nonlinear term must be important to balance the
small dissipative forcing, if the state of resonant dissipation is to be
attained. In the following we postulate that this resonance happens at a
quartic order. By inspection, one finds that only resonance conditions
involving the stiff modes can be satisfied in the atomic magnitude. In the
following we study the consequences that along some special circular orbits
such balancing mechanism is established by the existence of a quartic
resonant constant of motion. To discuss this stability by resonance we need
some results of the tangent dynamics along the orbital plane. This more
elaborate tangent dynamics is derived in a way analogous to Section V and
shall be given elsewhere, here we give only the main results. The
stiff-limit for the equal-mass two-body problem with retarded and advanced
fields is studied in Ref. \cite{astar2B}, and in the following we give the
generalization of these results for the case of arbitrary masses.

Up to linear order, the tangent dynamics along the orbital plane is
decoupled from the $z$-dynamics of Section V. To study this planar tangent
dynamics, it is convenient to describe the orbit along the $z=0$ plane using
gyroscopic coordinates%
\begin{eqnarray}
x_{k}+iy_{k} &\equiv &r_{b}\exp (i\Omega t)[d_{k}+\eta _{k}],
\label{gyroscopic} \\
x_{k}-iy_{k} &\equiv &r_{b}\exp (-i\Omega t)[d_{k}+\xi _{k}],  \notag
\end{eqnarray}%
where $\eta _{k}$ and $\xi _{k}$ are complex numbers defining the
perturbation of the circularity and the $d_{k}$ are defined below Eq. (\ref%
{Zperturb}). Because $x_{k}$ and $y_{k}$ are real, we should have $\eta
_{k}=\xi _{k}^{\ast }$ but a convenient way to minimize the quadratic
functional of Fokker's action is to treat $\eta _{k}$ and $\xi _{k}$ as
independent functions. To fix ideas we start from the stability of the
isolated two-body system, and again we define the normal-mode eigenvalue by $%
\lambda \Omega /\theta $, i.e., every coordinate perturbation oscillates in
time as $\exp (\lambda \Omega t/\theta )$ ( $\lambda $ is an arbitrary
complex number). The limiting form of the planar characteristic equation for
the isolated different-mass case is

\begin{equation}
(\frac{\mu \theta ^{4}}{M})\cosh ^{2}(\lambda )=1,  \label{Istar}
\end{equation}%
where $\mu $ is the reduced mass and $M\equiv m_{1}+m_{2}$ (for the
equal-mass case, our general Eq. (\ref{Istar}) reduces to Eq. (15) of Ref. 
\cite{astar2B}\ ). Along circular orbits both the planar and the
perpendicular linearized equations share the same limiting characteristic
Eq. (\ref{Istar}), as can be checked with Eq. (\ref{fourthZ}). For hydrogen $%
(\mu /M)$ is a small factor of about $(1/1824)$. It is important to
understand the structure of the roots of Eq. (\ref{Istar}) in the complex $%
\lambda $ plane, specially for $\theta $ of the order of the fine structure
constant. The very small parameter $\frac{\mu \theta ^{4}}{M}\sim 10^{-13}$
multiplying the squared hyperbolic cosine on the left-hand side of Eq. (\ref%
{Istar}) determines that $\sigma \equiv |\func{Re}(\lambda )|$ $\simeq \ln (%
\sqrt{\frac{4M}{\mu \theta ^{4}}})$. For the first $13$ excited states of
hydrogen this $\sigma $ is in the interval $14.2<|\sigma |<18.2$. The
imaginary part of $\lambda $ can be an arbitrarily large multiple of $\pi $,
such that the general solution to Eq. (\ref{Istar}) is 
\begin{equation}
\lambda =\pm (\sigma +i\pi q),  \label{unperastar}
\end{equation}%
where $q$ is an arbitrary integer. \ The plus or minus sign of Eq. (\ref%
{unperastar}) is related to the time-reversibility of the isolated two-body
system, a symmetry that is broken by radiation. This same exact phenomenon
happens for the $z$ direction. Next we include the dissipation of the DFS,
i.e., the Lorentz-Dirac self-interaction, a calculation performed by adding
the self-interaction force to the equations of motion of the isolated
system. Here we give only the characteristic planar equation up to $%
O(1/\lambda ^{4})$ 
\begin{equation}
(1+\frac{7}{\lambda ^{2}}+\frac{5}{\lambda ^{4}})(\frac{\mu \theta ^{4}}{M}%
)\cosh ^{2}(\lambda )=1-\frac{2}{3}\theta ^{2}\lambda +\frac{1}{9}\theta
^{4}\lambda ^{2}+(\frac{1}{\lambda }+\frac{5}{\lambda ^{3}})(\frac{\mu
\theta ^{4}}{M})\sinh (2\lambda )+...  \label{fourthXY}
\end{equation}%
It is remarkable that Eqs. (\ref{fourthXY}) and (\ref{fourthZ}) differ only
at the terms of $O(1/\lambda )$ and at the terms of type $\theta ^{4}\lambda
^{2}$, which describe small corrections for $\sigma $ in the atomic range.
The linear term on the right-hand side of Eqs. (\ref{fourthZ}) and (\ref%
{fourthXY}) with the $2/3$ coefficient is due to the self-interaction force.
This dissipative term breaks the time-reversal symmetry of Eq. (\ref{Istar}%
), and the roots of Eqs. (\ref{fourthZ}) and (\ref{fourthXY}) no longer come
in plus or minus pairs. Let $\lambda _{xy}$ be a root of \ Eq. (\ref%
{fourthXY}) with positive real part and $\ \lambda _{z}$ be a root of Eq. (%
\ref{fourthZ}) with a negative real part. In the stiff limit these are both
near one of the limiting roots (\ref{unperastar}) and can be expressed as%
\begin{eqnarray}
\lambda _{xy} &\equiv &(\sigma +\pi qi+i\epsilon _{1}),  \label{pair} \\
\lambda _{z} &\equiv &-(\sigma +\pi qi+i\epsilon _{2}),  \notag
\end{eqnarray}%
where the small perturbations $\epsilon _{1}$ and $\epsilon _{2}$ are so far
two arbitrary complex numbers. The second order balancing process studied
here involves the interaction of a $z$ mode with a planar mode, in the same
way used in Refs. \cite{PRL,normalDeluca}. This is because if the atom is to
recoil like a rigid body, one expects the fast dynamics to encircle the
circular orbit with fast spinning motions of balanced amplitude.

We henceforth assume heuristically that the state of resonant dissipation is
formed in a bifurcation involving perturbations along two special linear
modes of the tangent dynamics. We take a perpendicular normal mode of Eq. (%
\ref{fourthZ}) and a planar normal mode of Eq. (\ref{fourthXY}), with
eigenvalues $\lambda _{z}$ and $\lambda _{xy}$ respectively. The coordinate
of the planar normal mode is a linear combination of the four $\eta \xi $
gyroscopic coordinates: $u\equiv a_{1k}\eta _{k}+b_{1k}\xi _{k}$, while the
coordinate of the perpendicular $z$ normal mode is $Z\equiv
b_{1}z_{1}+b_{2}z_{2}$. Using the normal mode conditions $\theta \dot{u}%
=\Omega \lambda _{xy}u$ and $\theta \dot{Z}=\Omega \lambda _{z}Z$ one can
show that the quadratic form $uZ$ is a complex amplitude that oscillates
harmonically with the beat frequency $(\lambda _{xy}+\lambda _{z})\Omega
/\theta =i(\epsilon _{1}-\epsilon _{2})\Omega /\theta $. Our resonance
condition is to choose these two eigenvalues such that

\begin{equation}
\func{Re}(\lambda _{xy}+\lambda _{z})=0.  \label{necessary}
\end{equation}%
Condition (\ref{necessary}) avoids that the modulus of the amplitude $uZ$
has an exponential growth. We shall see that condition (\ref{necessary}) is
satisfied only for special discrete values of \ $\theta $. Since condition (%
\ref{necessary}) must be satisfied, we henceforth assume that $\epsilon _{1}$
and $\epsilon _{2}$ are real numbers, as any excess real part in Eq. (\ref%
{pair}) can be absorbed in the definition of $\sigma $. Condition (\ref%
{necessary}) is also the necessary condition to construct a resonant
constant in the neighborhood of the circular orbit; Because Fokker's action
is real, $\lambda _{z}^{\ast }$ and $\lambda _{xy}^{\ast }$ are also
eigenvalues to Eqs.(\ref{fourthZ}) and (\ref{fourthXY}) respectively, with
complex conjugate normal mode coordinates. Condition (\ref{necessary}) then
implies the usual necessary condition for a resonant constant 
\begin{equation}
\lambda _{xy}+\lambda _{z}+\lambda _{xy}^{\ast }+\lambda _{z}^{\ast }=0,
\label{furta}
\end{equation}%
as discussed in Refs. \cite{Furta, normalDeluca}. Using these complex
conjugate normal-mode coordinates and Eq. (\ref{necessary}), one can show
that the following quartic form is a constant of the motion up to higher
order terms \cite{Furta, normalDeluca}: 
\begin{equation}
C\equiv |u|^{2}|Z|^{2}+...  \label{normalform}
\end{equation}%
The quartic function of Eq. (\ref{normalform}) is constant because it is the
squared modulus of the harmonic amplitude $uZ=\sqrt{C}\exp (i(\epsilon
_{1}-\epsilon _{2})\Omega t/\theta )$. This necessary condition and the
continuation of the leading term (\ref{normalform}) to an asymptotic series
is discussed in Ref. \cite{normalDeluca}.\ 

The root-searching problem of Eq. (\ref{necessary}) is well posed and for
each integer $q$ conditions (\ref{fourthZ}) and (\ref{fourthXY}) together
with Eq. (\ref{pair}) determine a unique $\theta $ as a function of $q$ ,
i.e., $\theta $ is quantized by the integer $q$ that appears naturally in
Eq. (\ref{pair}). An asymptotic solution to condition (\ref{pair}) can be
obtained by expanding Eqs. (\ref{fourthZ}) and (\ref{fourthXY}) up to
quadratic order in $\epsilon _{1}$ and $\epsilon _{2}$ while treating $%
\sigma $ as an approximate constant. This approximation determines the
following discrete values for $\theta $

\begin{equation}
\theta ^{2}=\frac{6(\pi ^{2}q^{2}-\sigma ^{2})}{\sigma (\pi ^{2}q^{2}+\sigma
^{2})^{2}},  \label{asintheta}
\end{equation}%
and 
\begin{equation}
(\epsilon _{1}-\epsilon _{2})=\frac{4\pi q(3\sigma ^{2}-\pi ^{2}q^{2})}{%
\sigma (\sigma ^{2}+\pi ^{2}q^{2})^{2}}.  \label{e1e2}
\end{equation}%
According to QED, the circular Bohr orbits have maximal angular momenta for
that quantum number and a radiative selection rule ( $\Delta l=\pm \hbar $)
restricts the decay from level $k+1$ to level $k$ only, i.e. circular orbits
emit the first line of each spectroscopic series (Lyman, Balmer,
Ritz-Paschen, Brackett, etc...), henceforth called the QED circular line. We
have solved Eqs. (\ref{fourthZ}), (\ref{fourthXY}) and (\ref{pair}) with a
Newton method in the complex $\lambda $ plane. Every angular momentum $%
l_{z}=\theta ^{-1}$ determined by Eq. (\ref{necessary}) has a value in the
correct atomic magnitude ( $\theta ^{-1}\gtrsim 137.0$\ ); The first
resonance appears at $q=5$ for $\theta ^{-1}=252.4$ and the minimum value $%
\theta ^{-1}=48.52$ \ is attained at $q=7$, then $\theta ^{-1}$ increases
monotonically with $q$. The subset of Table 1 has frequencies $w_{DF}$
surprisingly close to the QED lines. These lines are for $q$ approximately
equal to an integer multiple of the integer part of $2\sigma $. We
conjecture here that among the resonances satisfying the necessary condition
(\ref{necessary}), only some have $|u|^{2}$ depending on the
translation-invariant quantities $(\xi _{1}-\xi _{2})$ and $(\eta _{1}-\eta
_{2})$ to allow a recoiling translation \cite{to be published}. In our
description the emission mechanism is at a frequency equal to the orbital
frequency $\Omega $ corrected by the frequency of the complex amplitude $uZ$
defined above Eq. (\ref{normalform}), as we explain below. The numerically
calculated angular momenta $l_{z}=\theta ^{-1}$ for this select subset are
given in Table 1, along with the orbital frequency in atomic units $%
(137^{3}\Omega )/\mu =(137\theta )^{3}$, the QED first frequency of the
series in atomic units $w_{QED}\equiv \frac{1}{2}(\frac{1}{k^{2}}-\frac{1}{%
(k+1)^{2}})$, and the frequency predicted by the dissipative Fokker model $%
w_{DF}\equiv (137\theta )^{3}+137^{3}\theta ^{2}(\epsilon _{1}-\epsilon
_{2}) $. We list only the first $13$ lines, which are the experimentally
observable, but we tested the agreement of the numerical calculations of the
Newton method with up to the $40^{th}$ circular line predicted by QED.
Beyond that, the asymptotic formula (\ref{asintheta}) shows that the
agreement is essentially for any integer $k$ because substitution of $%
q=[2\sigma ]k$ into Eq. (\ref{asintheta}) yields 
\begin{equation}
\theta ^{-1}=\sqrt{\frac{2\pi ^{2}}{3}}\sigma ^{3/2}k\sim 137.9k,
\label{hbar}
\end{equation}%
to be compared with the $137.036$ of QED. The agreement for any integer $k$
suggests that Eqs. (\ref{fourthZ}) and (\ref{fourthXY}) describe a linear
problem that is equivalent to Schroedinger's equation (linear operators with
the same spectrum are equivalent).\bigskip

\begin{tabular}{|l|l|l|l|}
\hline
$l_{z}=\theta ^{-1}$ & $(137\theta )^{3}$ & $w_{QED}$ & $w_{DF}$ \\ \hline
161.94 & 6.054$\times $10$^{-1}$ & 3.750$\times $10$^{-1}$ & 3.655$\times $10%
$^{-1}$ \\ \hline
283.52 & 1.128$\times $10$^{-1}$ & 6.944$\times $10$^{-2}$ & 6.774$\times $10%
$^{-2}$ \\ \hline
398.06 & 4.077$\times $10$^{-2}$ & 2.430$\times $10$^{-2}$ & 2.462$\times $10%
$^{-2}$ \\ \hline
520.29 & 1.826$\times $10$^{-2}$ & 1.125$\times $10$^{-2}$ & 1.110$\times $10%
$^{-2}$ \\ \hline
638.53 & 9.876$\times $10$^{-3}$ & 6.111$\times $10$^{-3}$ & 6.038$\times $10%
$^{-3}$ \\ \hline
752.27 & 6.039$\times $10$^{-3}$ & 3.685$\times $10$^{-3}$ & 3.710$\times $10%
$^{-3}$ \\ \hline
872.68 & 3.868$\times $10$^{-3}$ & 2.406$\times $10$^{-3}$ & 2.387$\times $10%
$^{-3}$ \\ \hline
988.16 & 2.664$\times $10$^{-3}$ & 1.640$\times $10$^{-3}$ & 1.650$\times $10%
$^{-3}$ \\ \hline
1110.15 & 1.879$\times $10$^{-3}$ & 1.173$\times $10$^{-3}$ & 1.168$\times $%
10$^{-3}$ \\ \hline
1226.95 & 1.392$\times $10$^{-3}$ & 8.678$\times $10$^{-4}$ & 8.677$\times $%
10$^{-4}$ \\ \hline
1344.30 & 1.058$\times $10$^{-3}$ & 6.600$\times $10$^{-4}$ & 6.615$\times $%
10$^{-4}$ \\ \hline
1462.14 & 8.226$\times $10$^{-4}$ & 5.136$\times $10$^{-4}$ & 5.153$\times $%
10$^{-4}$ \\ \hline
1580.44 & 6.513$\times $10$^{-4}$ & 4.076$\times $10$^{-4}$ & 4.090$\times $%
10$^{-4}$ \\ \hline
\end{tabular}%
\begin{tabular}{|l|}
\hline
$q$ \\ \hline
32 \\ \hline
55 \\ \hline
76 \\ \hline
98 \\ \hline
119 \\ \hline
139 \\ \hline
160 \\ \hline
180 \\ \hline
201 \\ \hline
221 \\ \hline
241 \\ \hline
261 \\ \hline
281 \\ \hline
\end{tabular}

\bigskip

Table 1: \ Numerically calculated angular momenta $l_{z}=\theta ^{-1}$ in
units of $e^{2}/c$, the orbital frequencies in atomic units $(137\theta
)^{3} $, the circular lines of QED in atomic units $w_{QED}\equiv \frac{1}{2}%
(\frac{1}{k^{2}}-\frac{1}{(k+1)^{2}})$ , the emission frequencies of the DFS
in atomic units $w_{DF}\equiv (137\theta )^{3}+137^{3}\theta ^{2}(\epsilon
_{1}-\epsilon _{2})$ and the values of the integer $q$ of Eq. (\ref{pair}).

\bigskip

In the DFS the interaction with a distant particle involves half the
retarded Li\'{e}nard-Wiechert potential plus half the advanced Li\'{e}%
nard-Wiechert potential (henceforth called the semi-sum). This semi-sum
yields a radiation magnetic field for the electron of (the far-magnetic
field) 
\begin{equation}
\mathbf{B}_{rad}=\frac{(\mathbf{a}_{-}\times \mathbf{\hat{n}}_{-})}{2(1-%
\mathbf{\hat{n}}_{-}\cdot \mathbf{v}_{-})^{2}r}-\frac{(\mathbf{a}_{+}\times 
\mathbf{\hat{n}}_{+})}{2(1+\mathbf{\hat{n}}_{+}\cdot \mathbf{v}_{+})^{2}r},
\label{Bfar}
\end{equation}%
where $\mathbf{v}$ and $\mathbf{a}$ are the electronic velocity and
acceleration, $\mathbf{\hat{n}}$ is a unit vector from the electron to the
observation point, the subindex minus sign indicates evaluation on the
retarded light-cone and the subindex plus sign indicates evaluation on the
advanced light-cone. These two light-cones are defined by $t_{\pm }=t\pm (r-%
\mathbf{\hat{n}}_{\pm }\cdot \mathbf{y})$, where $\mathbf{y}$ stands for the
electron's position. Along a precise circular orbit the first approximation
to Eq. (\ref{Bfar}) \ has a zero spatial average. \ For the next term we
avoid the Page expansion of Appendix A, because the deviating arguments are
large; We approximate the size of Eq. (\ref{Bfar}) by expanding the
denominators of \ Eq. (\ref{Bfar}), yielding the quadratic function%
\begin{equation}
\mathbf{B}_{rad}^{(1)}\simeq \frac{2(\mathbf{\hat{n}}\cdot \mathbf{v})(%
\mathbf{a}\times \mathbf{\hat{n}})}{r}.  \label{quadraticRad}
\end{equation}%
We can estimate $\mathbf{B}_{rad}^{(1)}$ of Eq. (\ref{quadraticRad}) by
noticing that along the $\mathbf{\hat{n}}_{\pm }=\mathbf{\hat{x}}$ direction
of the unperturbed plane this quadratic functional contains a product of the 
$z$ perturbed coordinate times the $x$ perturbed coordinate, i.e., the $u$
and $Z$ perturbations explained above Eq. (\ref{normalform}). Translating
the $u$ mode to Cartesian coordinates with Eq. (\ref{gyroscopic}) we obtain

\begin{equation}
\mathbf{B}_{rad}^{(1)}\propto \frac{2uZ}{r}\exp (i\Omega t).
\label{second_orderB}
\end{equation}%
According to Eq. (\ref{second_orderB}), the frequency of the emission line
is equal to $\Omega $ plus the frequency of the $uZ$ amplitude, 
\begin{equation}
w_{DF}=\Omega +(\epsilon _{1}-\epsilon _{2})\Omega /\theta ,
\label{emission}
\end{equation}%
with $\Omega $ given by Eq. (\ref{orbital}). Notice that the emitted
frequency of the DFS is naturally different from the orbital frequency. The
fact that the emission frequency of hydrogen is different from the orbital
frequency is a famous conundrum. The emission frequency of Eq. (\ref%
{emission}) contains differences of eigenvalues of the linear operator of
Eqs. (\ref{fourthXY}) and (\ref{fourthZ}) and is strikingly similar to the
Rydberg-Ritz combinatorial principle of quantum mechanics for the emission
lines.

\bigskip

\section{Conclusions and Discussion}

\bigskip

In the limit where the proton has an infinite mass, the concept of resonant
dissipation loses meaning because the center-of-mass coordinate no longer
plays a dynamical role. In this singular limit, there is a Lorentz frame
where the proton rests at the origin at all times, and the field at the
electron reduces to a simple Coulomb field in the DFS. The two-body dynamics
in the DFS reduces then to the dynamical system of Eliezer's theorem;
self-interaction plus a Coulomb field acting on the electron \cite%
{Eliezer,Parrott}. We repeat this correct dynamics because it is very
unpopular \cite{Eliezer,Parrott, Andrea, Massimo}; With inclusion of
self-interaction, it is impossible for the electron to "spiral into the
proton". Neither bound states nor dives are possible, only scattering states
exist. This result is in surprising agreement with our formula (\ref{hbar})
for the quantized angular momenta; If the mass of the proton is set infinite
in Eq. (\ref{hbar}), the quantized angular momenta become infinite
logarithmically, $\theta $ goes to zero, and the particles are unbound at an
infinite distance! One accomplishment of the present work is to recognize
that only the two-body problem can produce a physically sensible
electromagnetic-like model. Even though there is a dependence on the mass in
Eq. (\ref{hbar}), the logarithm of the mass ratio times $\theta ^{4}$ makes
the theory very insensitive to this mass ratio, such that the deuterium and
the muonium have essentially the same quantized angular momenta, in
reasonable agreement with QED. Qualitative disagreement would need an
exponentially massive charged particle. Fortunately to our present theory,
such particle does not exist in nature.

Another qualitative dynamical picture is suggested by Eliezer's result \cite%
{Eliezer,Parrott}; The dynamical phenomenon that the electron always turns
away from the proton along unidimensional orbits suggests that colinear
orbits are the natural attractors of the dissipative dynamics (a ground
state with zero angular momentum!). Along such orbits, the heavy particle
(the proton) moves in a non-Coulombian way and the self-interaction provides
the repulsive mechanism that avoids the collision at the origin. This is
again in agreement with the Schroedinger theory, where the ground state has
a zero angular momentum. Again, the infinite-mass case produces unphysical
dynamics; the electron turns away but then it runs away \cite{Parrott}. It
remains to be researched if the two-body case has a physical orbit for
zero-angular momentum orbits.

The theory of normal forms for delay equations is studied in Ref. \cite%
{normaldelay}. An analogous mathematical phenomenon is the
finite-dimensional center-manifold for equations with advance and delay
studied in connection with discrete shocks in the conservation laws of Refs. 
\cite{Majda, Gavage}. These conservation laws are similar to Dirac's
relativistic Schroedinger's equation, and this would be a natural bridge to
QED. Detailed construction of the resonant normal form is also needed to
discuss the width of the emission lines. In the dynamical process of
resonant exchange, the sharp line is emitted while the dynamics is locked to
the neighborhood of the resonant orbit, which according to QED is a
life-time of about $10^{6}$ turns in the hydrogen atom ($10^{-9}$ seconds).
We conjecture that when the metastable orbit breaks down, the dynamics falls
into the next metastable attracting orbit; another circular orbit, or into
the ground state \cite{to be published}.

The stiff modes of Eq. (\ref{pair}) introduce a fast (stiff) time-scale with
a frequency of the order of $\sigma /\theta \simeq 1400$ times the orbital
frequency, such that the time for a stiff jump of the dynamics is $\frac{1}{%
1400}$ times the orbital period, or $10^{-18}$ seconds! After this fast
timescale the resonance essentially locks the dynamics to the neighborhood
of the metastable resonant orbit. The fact that the equations of
electrodynamics describe stiff jumps in the phase space is largely
unexplored in the light of modern applied mathematics; mainly due to the
complexity involved. The dynamics starting from an asymptotic resonant orbit
to another of a neighboring $q$ is certainly described by a stiff jump, as
expected generically from any stiff equation. In Ref \cite{Grasman}, the
much simpler Van der Pol oscillator is worked out in detail as an example of
an equation of Lienard type that exhibits stiff jumps. In quantum mechanics
one seems to need the problematic concept of\ \ an "instantaneous quantum
jump", \ to describe the stiff passage from one quantum state to another. It
appears that classical electrodynamics prescribes exactly this qualitative
phenomenon; a quasi-instantaneous fast dynamics.

The dynamics in the DFS solves several conundrums of the classical hydrogen
atom and is similar to QED in many ways; (i) the radiated frequency is not
equal to the orbital frequency ( it is lesser than the orbital frequency,
see Table 1). (ii) the resonant orbits are naturally quantized by integers
and the radiated frequencies agree with the Bohr circular lines within one
percent average deviation. (iii) The ratio of the emitted frequency to the
orbital frequency is in reasonable agreement with QED. (iv) the angular
momenta of the resonant orbits are naturally quantized with the correct
Planck's constant. (v) the stability analysis uses a linear dynamical system
with delay, a dynamical system that needs an initial function as the initial
condition, just like Schroedinger's equation. The emitted frequencies are
then given by a difference of two eigenvalues of this linear operator, like
the Rydberg-Ritz combinatorial principle of quantum physics. (vi) The
eigenvalues of our linear operator have a large magnitude that does not
appear in the frequency. This large magnitude is given by a logarithm, just
like in the divergent perturbation theory for the Lamb-shift of QED. \ 

Recognizing the correct qualitative dynamics with the concept of resonant
dissipation has taken us very far; the stability analysis indicated the need
for resonances, and these turned out to be satisfied only for the stiff
modes and precisely in the atomic magnitude! The stiff modes also provide a
natural integer to label the resonant orbits. We selected the values of $q$
among the larger set predicted by the necessary condition (\ref{necessary}),
showing that Eq. (\ref{necessary}) is not in disagreement with QED. A
sufficient condition should be part of the extra work to understand the
unfolding of the bifurcation leading to the state of resonant dissipation.
The large body of qualitative and quantitative agreement suggests that an
extensive study of electromagnetic-like models \cite{EliezerReview}, of
which the DFS is only a generic example, could offer an explanation of QED
in terms of a stiff dynamical system with third derivatives and delay.

\section{Acknowledgements:}

I thank L. Galgani, A. Carati, R. Napolitano, S. Ruffo and A. Lichtenberg
for the support during the many years of this research. I also thank A.
Ponno, M. Marino, A. Staruszkiewicz , \ A. Piza, \ S. Rodrigues, H. Von
Baeyer, F. Alcaraz and S. Mizrahi for discussions\bigskip .

\bigskip

\section{Appendix A: Physical justification of the DFS}

In Dirac's theory \cite{Dirac} the self-interaction is given by the
sourceless combination of half of the retarded Li\'{e}nard-Wiechert
self-potential minus half of the advanced Li\'{e}nard-Wiechert
self-potential, i.e., the semi-difference \cite{Dirac}. This gives the
following concise description of the DFS; Charges interact with themselves
via the semi-difference of Li\'{e}nard-Wiechert self-potentials and with
other charges via the semi-sum of Li\'{e}nard-Wiechert potentials. In the
following we try to fit our ad-hoc DFS into Dirac's theory as an effect of
the physical boundaries on the fields. Dirac's electrodynamics of point
charges \cite{Dirac} uses the retarded potential $F_{\mu k,ret}^{\nu }$
produced by each particle $k$ and an incident free field $F_{\mu ,in}^{\nu }$%
. In Dirac's theory the electron and the proton of a hydrogen atom have the
following equations of motion \cite{Dirac}%
\begin{eqnarray}
m_{1}\dot{v}_{1\mu }-\frac{2}{3}\ddot{v}_{1\mu }-\frac{2}{3}%
||v_{1}||^{2}v_{1\mu } &=&-(F_{\mu ,in}^{\nu }+F_{\mu 2,ret}^{\nu })v_{1\nu
},  \label{Diracmotion} \\
m_{2}\dot{v}_{2\mu }-\frac{2}{3}\ddot{v}_{2\mu }-\frac{2}{3}%
||v_{2}||^{2}v_{2\mu } &=&(F_{\mu ,in}^{\nu }+F_{\mu 1,ret}^{\nu })v_{2\nu },
\notag
\end{eqnarray}%
where double bars stand for the Minkowski scalar product, the electron and
the proton have charges $-1$ and $1$ respectively and the speed of light is $%
c=1$. Since the DFS uses the semi-sum instead of the retarded-only
potential, from the perspective of Dirac's theory this demands the following
constraints on the free field $F_{\mu ,in}^{\nu }$ 
\begin{eqnarray}
F_{\mu ,in}^{\nu }(x_{1}(t)) &=&\frac{1}{2}[F_{\mu 2,adv}^{\nu
}(x_{1}(t))-F_{\mu 2,ret}^{\nu }(x_{1}(t))],  \label{constr1} \\
F_{\mu ,in}^{\nu }(x_{2}(t)) &=&\frac{1}{2}[F_{\mu 1,adv}^{\nu
}(x_{2}(t))-F_{\mu 1,ret}^{\nu }(x_{2}(t))],  \label{constr2}
\end{eqnarray}%
where the field of each particle is to be evaluated along the trajectory of
the other particle, as indicated by the parenthesis after each field. Since
both the advanced and the retarded fields satisfy Maxwell's equations, the
semi-difference is a free field, as assumed. The incident wave can be
generated by the boundary conditions on the fields. For example, the
reflections of the radiation by other atoms of a diluted gas could play the
role of such a boundary condition.

The semi-difference evaluated at the particle itself is the familiar
self-interaction of the Dirac theory \cite{Dirac}, and Eqs. (\ref{constr1})
and (\ref{constr2}) have instead the semi-difference evaluated at the
position of the other particle. Using the Page expansion of the Li\'{e}%
nard-Wiechert fields, we find that the electric field of this
semi-difference is approximated by the third derivative of the other
particle's coordinate, as discussed in Refs. \cite{PRL,normalDeluca}. In
this approximation with the Page series \cite{PRL,normalDeluca}, the
incident electric field evaluated at the proton, Eq.(\ref{constr2}), is 
\begin{equation}
E(x_{2}(t))\simeq \frac{2}{3}\dddot{x}_{1}.  \label{electric}
\end{equation}%
Along the unperturbed orbit of Fig.1, Eq. (\ref{electric}) is an
electromagnetic field rotating at the orbital frequency. For orbits in the
atomic magnitude the electric field of Eq. (\ref{electric}) has an intensity
that turns out to be of the order of the polarized vacuum of QED, as
discussed in Ref.\cite{Dalibard}. This shows that our needed homogeneous
field has the correct physical magnitude of the QED vacuum polarized by the
hydrogen atom. We see that the ad-hoc DFS demands a free field produced by
the boundaries that is calculated to have a physically sensible order of
magnitude. This approach to justify the DFS with a free field produced by
the boundaries is similar to that of the stochastic electrodynamics of Refs. 
\cite{Marshall, Boyer}.

Finally, we mention a more radical alternative to justify our ad-hoc DFS, by
generalizing Dirac's theory such that the DFS would be derived \emph{from
principle}. This approach was taken by Eliezer and this generalization,
henceforth called the Eliezer's setting (ES), is discussed in the excellent
review of Ref. \cite{EliezerReview}. The ES involves the advanced
interactions naturally, exactly in the same form of the DFS! Better still;
the ES \cite{EliezerReview} contains an arbitrary parameter, and it would be
highly desirable to experiment with stability analysis and the concept of
resonant dissipation in the ES \cite{EliezerReview}. Even though the ES
involves delay, advance and third derivatives exactly like the DFS, the
coefficients in the ES are never equal to those of the DFS. Our preliminary
findings with the DFS\ suggest a future for this enterprise in the
qualitative behavior of electromagnetic-like dynamics, one that could
describe QED by a stiff dynamical system with delay.

\bigskip

\bigskip

\section{Appendix B: Darwin and the soft Coulombian modes}

In this appendix we calculate $b_{1}$ and $b_{2}$ of Eq. (\ref{defradius})
as a function of $m_{1}$, $m_{2}$ and $\theta $. The radial component of the
electron's equation of motion along the circular orbit is \cite{Schild}%
\begin{equation}
\frac{m_{1}b_{1}r_{b}\theta ^{2}}{\sqrt{1-\theta ^{2}b_{1}^{2}}}=\frac{1}{%
S^{3}}\{[C^{2}+\theta ^{2}S(S-1)](b_{1}+b_{2}\cos (\theta ))+\theta S(\sin
(\theta )-\theta \cos (\theta ))b_{2}\},  \label{radialEq1}
\end{equation}%
where $C$ and $S$ are defined in Eqs. (\ref{defC}) and (\ref{defS})
respectively. Our Eq. (\ref{radialEq1}) is Eq. (3.2) of Ref. \cite{Schild}
after use of Eq. (\ref{circularcone}) and the identity%
\begin{equation}
(1-\theta ^{2}b_{1}^{2})(1-\theta ^{2}b_{2}^{2})=C^{2}+\theta ^{2}(S-2)S.
\label{magic}
\end{equation}%
The radial equation for the proton is obtained by exchanging the subindices $%
1$ and $2$ in Eq. (\ref{radialEq1}). There are three equations involving $%
b_{1}$, $b_{2}$ , $\theta $ and $r_{b}$; \ (i) \ Eq. (\ref{radialEq1}), (ii)
the equation for the proton, obtained by exchanging indices $1$ and $2$ in
Eq. (\ref{radialEq1}), and (iii) the light-cone condition, Eq. (\ref%
{circularcone}) 
\begin{eqnarray}
\frac{m_{1}b_{1}r_{b}\theta ^{2}}{\sqrt{1-\theta ^{2}b_{1}^{2}}} &=&\frac{1}{%
S^{3}}\{[C^{2}+\theta ^{2}S(S-1)](b_{1}+b_{2}\cos (\theta ))+\theta S(\sin
(\theta )-\theta \cos (\theta ))b_{2}\},  \notag \\
\frac{m_{2}b_{2}r_{b}\theta ^{2}}{\sqrt{1-\theta ^{2}b_{2}^{2}}} &=&\frac{1}{%
S^{3}}\{[C^{2}+\theta ^{2}S(S-1)](b_{2}+b_{1}\cos (\theta ))+\theta S(\sin
(\theta )-\theta \cos (\theta ))b_{1}\},  \notag \\
b_{1}^{2}+b_{2}^{2}+2b_{1}b_{2}\cos (\theta ) &=&1.  \label{AP1}
\end{eqnarray}%
For small values of $\theta $ (atomic physics), we can \ solve Eqs. (\ref%
{AP1}) in a power series of $\theta $ with a symbolic manipulation software,
yielding%
\begin{eqnarray}
b_{1} &=&\frac{m_{2}}{M}(1+\frac{\mu \theta ^{2}}{2M})+\theta
^{4}D(m_{1},m_{2})+...  \label{fourthbexp} \\
b_{2} &=&\frac{m_{1}}{M}(1+\frac{\mu \theta ^{2}}{2M})+\theta
^{4}D(m_{2},m_{1})+...,  \notag
\end{eqnarray}%
where%
\begin{equation}
D(m_{1},m_{2})\equiv (\frac{\mu }{24M})[\frac{%
12m_{1}^{3}-13m_{2}^{3}-5m_{1}m_{2}^{2}+11m_{2}m_{1}^{2}}{M^{3}}].
\label{defD}
\end{equation}%
It is easy to continue this power series, but for the stiff limit in the
atomic magnitude, even the $\theta ^{2}$ correction already gives a very
small correction. The orbital frequency is determined by 
\begin{equation}
\Omega =\frac{v_{1}}{b_{1}r_{b}}=\mu \theta ^{3}[1+(\frac{1}{2}+\frac{\mu }{%
2M})\theta ^{2}+...],  \label{orbital}
\end{equation}%
the first term is Kepler's third law if we use $\theta =\Omega r_{b}$, and
the next term is the Darwin correction. More information about the isolated
two-body problem can be found in Refs. \cite{Hans,astar2B}.

As an application of the above expansion, we calculate the soft Coulombian
modes of Eq. (\ref{detZ}) at a finite $p$ by expanding up to $O(\theta ^{3})$
:%
\begin{equation}
\det Z=\frac{M}{\mu }p^{2}(1+p^{2})[1-\frac{\theta ^{2}}{2}(1-\frac{12\mu }{M%
})]-\frac{2}{3}\frac{M}{\mu }p^{3}(p^{2}+2\frac{\mu }{M}).  \label{soft}
\end{equation}%
The soft roots of $\det Z=0$ for Eq. (\ref{soft}) are the Galilean
translation mode $p=0$ (a double root) and the oscillatory solutions $%
p\simeq \pm i$ \ that have a real part describing the radiative damping of
the DFS, a familiar feature. We had partial success describing the atomic
dynamics of helium with the Darwin approximation \cite{discrete}, and the
tools of stability analysis used here were already used in Refs. \cite%
{PRL,normalDeluca}. The concept of \ resonant dissipation is new, and it is
a generalization of the concept of a non-ionizing dynamics of Ref.\cite%
{discrete}. Unfortunately, the theory of Refs. \cite{PRL,normalDeluca,
discrete} fails to describe discrete states for hydrogen because the soft
Darwin modes are neutrally stable. As we have seen here, it is the stiff
modes that equilibrate the dynamics, and those are beyond the Darwin
approximation.

\bigskip

\section{Captions}

\bigskip

Fig. 1: \ The unperturbed circular orbit with the particles in diametral
opposition at the same time in the inertial frame. Indicated is also the
advanced position of particle $2$ and the angle travelled during the
light-cone time. The drawing is not on scale; The circular orbit of the
proton has an exaggerated radius for illustrative purposes. Arbitrary units.

\end{document}